\documentclass{aa}
\usepackage{psfig}
\usepackage{graphicx}
\def\sax{{\em Beppo}SAX}
\def\grb{GRB980519}
\def\deg{\ifmmode^{\circ}\else$^{\circ}$\fi} % overwrite \deg in LaTeX
\def\farcm{\hbox{$\,.\!\!^{\prime}$}}
\def\farcs{\hbox{$\,.\!\!^{\prime\prime}$}}
\def\fs{\hbox{$\,.\!\!^{\rm s}$}}

\begin{document}

\title{The X-ray afterglow of \grb}

\author{L. Nicastro \inst{1}
\and L. Amati \inst{2}
\and L. A. Antonelli \inst{3,4}
\and E. Costa \inst{5}
\and G. Cusumano \inst{1}
\and M. Feroci \inst{5}
\and F. Frontera \inst{2,6}
\and E.~Palazzi \inst{2}
\and E.~Pian \inst{2}
\and L. Piro \inst{5}
}

\institute{
 {Istituto di Fisica Cosmica con Applicazioni all'Informatica, CNR,
 Via U. La Malfa 153, I-90146 Palermo, Italy}
\and %2
 {Istituto Tecnologie e Studio Radiazione Extraterrestre, CNR, Via Gobetti 101,
 I-40129 Bologna, Italy}
\and %3
 {Osservatorio Astronomico di Roma, Via Frascati 33, I-00040 Monteporzio Catone
 (RM), Italy}
\and %4
 {BeppoSAX Scientific Data Center, Via Corcolle 19, I-00131 Roma, Italy}
\and %5
 {Istituto Astrofisica Spaziale, CNR, Via Fosso del Cavaliere, I-00131 Roma,
 Italy}
\and %6
 {Dipartimento di Fisica, Universit\`a di Ferrara, Via Paradiso 11,
 I-44100 Ferrara, Italy}
} 
\offprints{nicastro@ifcai.pa.cnr.it}
\date{Received ........; accepted .........}
\thesaurus{011 (13.07.1; 13.25.1)}

\maketitle
\markboth{L. Nicastro et al.: The X-ray afterglow of \grb}
         {L. Nicastro et al.: The X-ray afterglow of \grb}

\begin{abstract}
Over a total of 20 gamma-ray bursts localized with arcmin accuracies,
\grb\ represents the 13$^{\rm th}$ detected by the \sax\ Wide Field Cameras
(WFCs). An X-ray TOO observation performed by the \sax\ Narrow Field
Instruments (NFIs), starting about 9.5 hours after the high energy event,
revealed X-ray afterglow emission in the 0.1--10 keV energy range.
The flux decay was particularly fast with a power-law index of
$\simeq 1.8$.
This is the fastest decay so far measured. Signs of bursting activity are
evident. The power-law spectral index of $2.8^{+0.6}_{-0.5}$ is quite
soft but not unique among GRB afterglows.

$VR_cI_c$ optical emission was detected as soon as 8 hours after the GRB
and the power-law flux decay in all these bands were all consistent with
$\delta \simeq 2.0$. As for the X-ray, this is the fastest of all the
9 optically identified afterglows but GRB980326. A candidate host galaxy with
magnitude $R_c = 26$ has been reported and variable radio emission detected.


\end{abstract}

\section{Introduction}

\begin{figure}[ht]
\centerline{
\psfig{figure=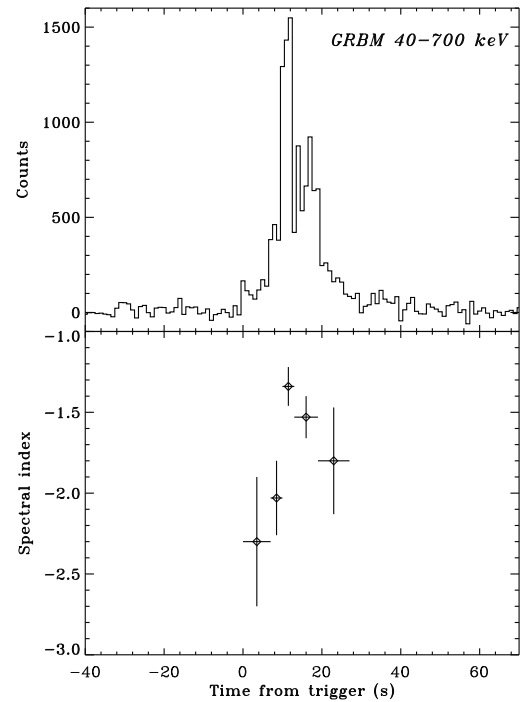,height=10cm}
}
\caption{
 The 1 s GRBM light curve of \grb\ and its power-law photon spectral index
 evolution.
}
\label{grbmlc}
\end{figure}

\grb\ was detected on 1998 May 19, 12:20:13 UT by CGRO--BATSE and
\sax--GRBM.  It was in the field of view of the \sax--WFC 2,
allowing an estimate of its position with a 3 arcmin error circle at
R.A. = 23$^{\rm h}$22$^{\rm m}$15$^{\rm s}$
       Dec. = $+77^\circ15\farcm 0$ (J2000).
The GRBM light curve (lasting $\simeq 30$ s) together with
the spectral evolution in the 40--700 keV range is shown in
Fig. \ref{grbmlc}. The soft--hard--soft evolution is evident in the plot.
In the 2--27 keV WFC band a similar behaviour is observed but
the light curve is much
more structured than the high energy one and it lasts for
$\simeq 180$ s; the emission starts
$\simeq 50$ s before the GRB trigger and stops $\simeq 130$ s later
(in 't Zand et al. 1998).
%In spite of the small number of GRBs simultaneously detected in X- and
%gamma-rays,
%such prominent X-ray activity, especially before the gamma-ray event,
%seems to be a rare event and its interpretation is completely open.

The average spectrum of the GRBM data can be well fitted with a single
power-law with a photon spectral index ($N_E\propto E^{-\alpha}$)
$\alpha=1.62\pm 0.10$ with $\chi^2\simeq 1$.
The 40--700 keV fluence (over 27 s) is
$F_\gamma = (8.1\pm 0.5)\times 10^{-6}$ erg cm$^{-2}$ and the hardness ratio
$f_{100-300/50-100} = 2.2 \pm 0.2$.
Description of the method adopted for GRBM spectra deconvolution is reported
by Amati et al. (this workshop).

\section{\sax\ NFIs observation and discussion}

A follow-up observation performed with the \sax\ NFIs started
less than 10 hrs after the trigger and a weak, rapidly decaying
X-ray source was detected (Nicastro et al. 1998a).
The decay was not monotonic,
but the low counting rate did not allow us to reconstruct a detailed
light curve.
%Figure \ref{mecsimg} shows the \sax--MECS 2+3 smoothed and scaled
%images accumulated in three time intervals. The WFC 3$'$ radius error circle
%is shown.

\begin{figure}[tb]
\centerline{
\psfig{figure=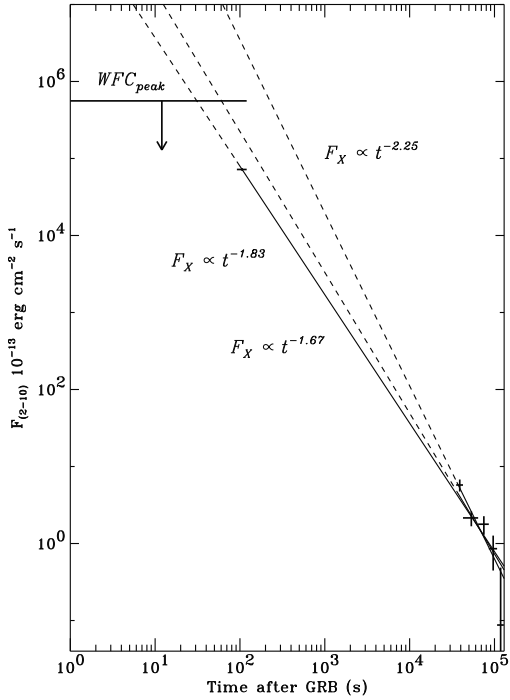,height=10cm}
}
\caption{
 2--10 keV X-ray flux decay as seen by the \sax\ MECS. The upper limit to the
 flux during the burst and the last WFC measured values are shown.
 Three possible fitted power-law decays are reported.
}
\label{xdecay}
\end{figure}

Figure \ref{xdecay} shows the 2--10 keV (MECS) flux decay and
three possible
power-law decay fits $F_x \propto t^{-\delta}$.
It can be seen that constraining the fit to connect with the last part
of the WFC detected flux, gives a {\em minimum} decay slope
$\delta = 1.67\pm 0.03$. On the other hand, if the first point
of the NFIs observation is {\em not} a peak superimposed to the monotonic
decay, then we have $\delta = 2.25\pm 0.22$. This can be considered
a {\em maximum} decay slope. It is realistic to suppose that the {\em real}
slope is close to $\delta = 1.83\pm 0.30$ obtained excluding the
first NFIs point from the fit. In any case, this is the most rapid decay
for all the 13 GRB afterglows detected so far, typical
values ranging between $1.1 \div 1.4$.

\begin{figure}
\centerline{
\psfig{figure=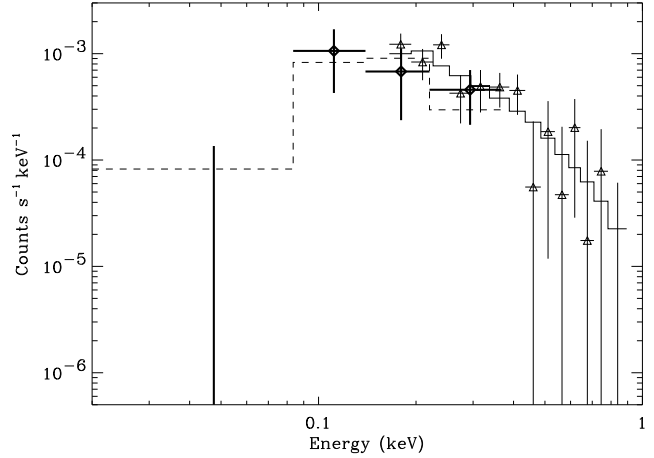,width=8.7cm}
}
\caption{
 LECS+MECS 0.1--10 keV spectrum of the GRB afterglow. It is
 $\alpha=2.8^{+0.6}_{-0.5}$ and $N_H\approx
 0.3-2\; \times 10^{22}$ cm$^{-2}$.
}
\label{toospe}
\end{figure}

Spectral fitting of the 0.1--10 keV LECS+MECS data using an absorbed
power-law gave
a quite soft photon index of $2.8^{+0.6}_{-0.5}$
and $N_H$ in the range
$0.3-2\; \times 10^{22}$ cm$^{-2}$ (see Fig. \ref{toospe}).
The 0.1--2 keV flux is
$(7.9\pm 2.5)\times 10^{-14}$ erg cm$^{-2}$ s$^{-1}$
 while the 2--10 keV flux is
$(1.4\pm 0.3)\times 10^{-13}$ erg cm$^{-2}$ s$^{-1}$.
Further details are given in Nicastro et al. (1998b).

Optical observations, started as early as 8 hours
after the GRB, resulted in the detection of the afterglow
(OT) in $VR_cI_c$ bands.
The power-law decays in all these bands were all consistent with
$\delta \simeq 2.0$.
%  As for the X-ray afterglow, this is the
%fastest decay measured.
Deep observations performed $\simeq 66$ days after the
burst with the 6-m telescope of the SAO--RAS
revealed that at the position of the OT there is a faint object,
possibly a galaxy, of magnitude $R_c \simeq 26$
(Sokolov et al. 1998b).
It is worth to note that for all 9 optically identified GRBs,
there are indications of the presence of an underlying host galaxy
(Hogg \& Fruchter 1998; Sokolov et al. 1998a).

The afterglow was also detected in the radio band by the VLA
(Frail et al. 1998) at
 R.A. = 23$^{\rm h}$22$^{\rm m}$$21\fs49$
 Dec. = $+77^\circ15'43\farcs 2$ (J2000, $\pm 0\fs1$).

\begin{acknowledgements}
This research is supported by the Italian Space Agency (ASI) and
Consiglio Nazionale delle Ricerche (CNR). BeppoSAX is a major program
of ASI with participation of the Netherlands Agency for Aerospace
Programs (NIVR). All authors warmly thank the extraordinary teams
of the BeppoSAX Scientific Operation Center and Operation Control
Center for their enthusiastic support to the GRB program.
K. H. is grateful to the US SAX Guest Investigator program for support.
\end{acknowledgements}

\end{document}